\documentstyle[floats,aps,epsf,rotate]{revtex}

\makeatletter
\def\preprint#1{%
\def\@preprint{\noindent\hfill\hbox{#1}\vskip 10pt}%
}
\makeatother
\def\BRA{\left\langle}
\def\m{{\bf m}}
\def\KET{\right\rangle}
\begin{document}
\preprint{to appear in Z. Phys. B}
\draft
\twocolumn[\hsize\textwidth\columnwidth\hsize\csname @twocolumnfalse\endcsname
\title{Lattice Electrons on a Cylinder Surface in the Presence of
Rational Magnetic Flux and Disorder\cite{goetze}} 
\author{Christian Schulze, J\'anos Hajdu, Bodo Huckestein, Martin Janssen}
\address{ Institut f\"ur Theoretische Physik, Universit\"at zu K\"oln, 
 Z\"ulpicher Str. 77, 50937 K\"oln,
 Germany}  
 
\date{Dec. 19., 1996}
\maketitle

\begin{abstract}
We consider a disordered two-dimensional
 system of independent lattice electrons in
a  perpendicular magnetic field with rigid confinement in one
direction and generalized periodic boundary conditions (GPBC) in the
other
direction. The objects investigated numerically are the orbits in the
plane spanned by the energy eigenvalues and the corresponding center
of mass coordinate in the confined direction, parameterized by the
phase characterizing the GPBC. The Kubo Hall conductivity is expressed
in terms of the winding numbers of these orbits.
For vanishing disorder the spectrum of the system consists of Harper
bands with energy levels corresponding to the edge states within the
band gaps. Disorder leads to broadening of the bands. For sufficiently
large systems localized states occur in the band tails. We find that
within the mobility gaps of bulk states the Diophantine equation
determines the value of the Hall conductivity as known for systems
with torus geometry (PBCs in both directions). Within the spectral
bands of extended states the Hall conductivity fluctuates
strongly. For sufficiently large systems the generic behavior of
localization-delocalization transitions characteristic for the quantum
Hall effect are recovered.
\end{abstract}

\vskip2pc]

\section{Introduction}
The two-dimensional electron gas  in the presence of a strong perpendicular
magnetic field provides a rich 
 variety of  physical phenomena.
The most prominent is the quantum Hall 
effect \cite{PraGirv,JanB}.
 Another fascinating  phenomenon is the self-similar energy spectrum
(Hofstadter's butterfly) brought about by a periodic  lattice
potential and the magnetic field    \cite{Hof76,Exp}.
These phenomena are due to two features originating from the magnetic
field:
 chirality and a characteristic length scale.
The chirality is due to the axial vector character
of the magnetic field and is responsible for the   Hall effect in the
presence of an electric field and current carrying 
 edge states in confined systems. The length scale is defined
 by the size of the cell associated with a single
 flux quantum and is called the magnetic length. 
In a periodic potential commensurability effects occur with respect to
the ratio of 
  lattice constant and  magnetic 
length.  
In the presence
of  disorder also the localization length comes into play which
measures
the spatial extension of wave functions.
The natural way to investigate the interplay of these different length scales 
is to study the (dissipative and Hall) conductances of multi-terminal
systems.  Recently, a two-terminal
conductance calculation for a clean system was performed by Skj{\aa}nes
{\em et
al.} \cite{Skj94} and a multi-terminal conductance calculation
has been performed by Aldea {\em et al.} \cite{Ald96}.
 For the time being the attempts of calculating
directly the multi-terminal
conductances 
of  disordered systems 
are limited by computational practicability. Therefore, in the present
paper we restrict ourselves to investigate 
the energy spectrum, its sensitivity to changing the  boundary conditions
 and the Kubo Hall
conductivity.

The system to be studied is a  
two-dimensional Anderson tight-binding model with
Peierls substitution (in Landau gauge)
  and generalized periodic
boundary conditions (GPBC) in one direction and  different 
  boundary conditions in the other direction.
This system  has been investigated by many authors and several results are
well established.
In the limit of vanishing disorder 
 the Schr\"odinger equation reduces to a
one-dimensional finite difference equation known as Harper's
equation which can be studied
effectively by transfer matrix methods.
Imposing  periodic boundary conditions
in both directions corresponds to a  torus. If one of these is
replaced by rigid wall confinement we are dealing with a cylinder.
For both geometries the  energy spectrum is periodic in the flux per
unit lattice cell (denoted by  $\alpha$) with period $1$. 
Concerning the torus  the following results are  established:
(i) In the case of commensurability, i.e.  
  rational values $\alpha=p/q$ 
the energy spectrum consists of $q$ Harper bands 
separated by gaps.
The gap structure  gives rise to  Hofstadter's butterfly \cite{Hof76}.
(ii) The energy spectrum is symmetric with respect to  zero  energy.
(iii) The Hall conductivity is antisymmetric with respect to zero
 energy  (referred to as particle-hole symmetry).
(iv) Within the band gaps the  Hall conductivity is quantized in
integer multiples of $e^2/h$; its value is given by a topological
quantum number, the first Chern number on the magnetic Brillouin zone. 
The integer  can be determined from the Diophantine equation \cite{Tho82}
\begin{equation}\label{1}
        n=pt_n -qs_n \, .
\end{equation}
Here $t_n,s_n$ are integers corresponding to the $n$-th gap
and the Hall conductivity there is  $\sigma_H=t_n(e^2/h)$.
The solution of Eq.~(\ref{1}) is to be taken under the requirement
$|t_n|\leq q/2$.

For the cylinder geometry the
energy spectrum  has also been shown to consist of  $q$ Harper bands
 but the gaps
are filled with states  localized near the edges in
the confined direction ($x$), and  extended in the other.
These edge states  carry a chiral current 
and give rise to the same quantization
of the Hall conductivity as  obtained for the torus geometry
\cite{Tou83}. 
Recently 
\cite{Hat93} 
an  interpretation of the quantization in terms of topological invariants
corresponding to these edge states has been proposed and related to the
Chern numbers  of the torus geometry. However, this relation could be
established for no disorder only (see also \cite{Str94}).

So far, most  studies have been focused on the  ideal  system 
where analytical results can be obtained (especially the Diophantine equation
for the Hall conductivity). It has been argued that disorder
will not change the quantized value of the  Hall conductivity
  as
long as the disorder does not close the corresponding gap. This is
certainly true
for the torus geometry where  the Hall conductivity is given by the
topologically  stable Chern number as long as the Fermi energy is within
an energy gap. The argument is however not proven 
for the cylinder geometry where no
such gaps  exist. Moreover, in the torus geometry disorder leads to
levels within the gaps, although the corresponding states are expected
 to be localized and so do not   change the quantization
as given by  the  Diophantine equation. 
For the torus geometry the following scenario is expected to occur:
On increasing the
disorder strength localized states appear
first in the tails of the disorder broadened Harper bands.
 When the disorder is  further increased 
the Harper bands begin to  overlap strongly 
and form new bands with possibly less regions of extended states.
Yet,  for  Fermi energies at which  a small density of 
localized states is situated in a previous  gap the topological
stability of the Chern number  still  determines  the Hall
conductivity via the Diophantine equation (cf.~Sec.~VIII E in\cite{Huc94}).

For the disordered cylinder model (cf.~\cite{Sch84,Aok84})
 we cannot rely on the topological
stability
of Chern numbers when discussing the Hall conductivity.
Therefore, in the present work we  take a different starting point: 
we investigate  the Hall conductivity in an appropriate version
for the cylinder geometry (cf.~\cite{Haj87})
and show that in this case also it has   a topological
meaning as it can be expressed in terms of  a winding number. This
 winding number 
has an intuitive 
interpretation: it is the number of oriented windings of  those orbits,
in a diagram of energy eigenvalues versus  the center of mass
coordinate, $X$, that connect opposite edge regions.
The orbits are
parameterized by a continuous quantum number which  labels
generalized periodic boundary conditions in the direction of current
flow ($y$-direction). 
This winding number is similar to the Chern number when interpreted as the
winding number of zeroes of the wavefunction with respect to the change of
boundary conditions \cite{Koh85,Aro88}.  
We  study numerically the influence of increasing disorder on the validity of 
the Diophantine  quantization. 
We also investigate the Hall conductivity  within the  Harper bands
and show  that the Hall conductivity strongly fluctuates as a
function of the Fermi energy. For weak disorder and finite system
 sizes
an unexpected
 collapse of the Hall conductivity occurs in the
centers of the bands.
This is, however,  no surprise when consulting our version of
the Kubo Hall conductivity: As long as states in the band centers are extended
and
do not display large spatial fluctuations, $X$ will be
concentrated at the center of the system. Then the Hall conductivity
stays close to zero.   
We expect  that
as soon as the disorder is strong enough in finite systems or
 arbitrary small but finite  in infinite systems  the collapse
will disappear and (apart from strong fluctuations) the 
Hall conductivity effectively 
interpolates between subsequent the  quantized values.
The reason for this expectation  is that extended states
 (which in large systems are
restricted to the energy band centers) show strong spatial fluctuations which
are extremely sensitive to the variation of boundary conditions, i.e.
 the center
coordinate $X$ of such states can be shifted to an  arbitrary position
within the system by an appropriate change in the boundary conditions.
To examine the spatial fluctuations of extended eigenstates and
the universal properties of the localization-delocalization
transition we  calculate the  scaling behavior
of the distribution of wave function amplitudes of
extended states.
This (multifractal) analysis (cf.~Chap.~12 of \cite{JanB}) 
confirms our expectation   and, furthermore, we
find reasonable
agreement with the  scaling exponents obtained 
for  other model systems of the integer quantum Hall effect.

The paper is organized as follows:
In Sec.~\ref{model} we  introduce the  model and in
 Sec.~\ref{topology} 
the methods of calculating  energy orbits  and the Hall
conductivity. The quantization of the Hall conductivity
is determined by  the topology of the energy orbits.
 In Sec.~\ref{orbitsweak} we  present numerical results 
for the energy orbits corresponding to ideal  systems with cylinder
geometry and, by comparing to the torus geometry, we demonstrate how edge
states
appear in the energy gaps of the bulk system.
 We then turn over to weak disorder and investigate its
influence
on the topology of energy orbits.
 The Hall conductivity is unaffected 
for Fermi energies situated in the bulk  energy gaps and is still
given by the Diophantine equation, Eq.~(\ref{1}).
For Fermi energies situated in the bulk  energy bands the Hall
conductivity
drops down and fluctuates around zero. To see whether 
this unexpected behavior is still valid for larger systems
with strong disorder we present calculations of energy orbits in
such systems in Sec.~\ref{orbitsstrong}.
We discuss the fluctuation properties
of the wave function for extended states in the weak and strong scattering
regimes, respectively, and we conclude
 that for the latter the amplitude fluctuations are strong and 
 display universal
characteristics of localization-delocalization transitions in quantum Hall
systems. 
 The Hall conductivity  interpolates between
adjacent
plateau values as soon as the system size becomes much larger than the
microscopic scales of the model but still exhibits strong mesoscopic
fluctuations. Our conclusions are summarized in Sec.~\ref{conclusion}.

\section{The model}\label{model}
Consider a two-dimensional (2D)
 square lattice with lattice constant $a$ and extensions 
$L_x=M_xa$, 
 $L_y=M_ya$. With a set of
 orthonormal  quantum states $\left|\m\right\rangle$ localized at
 lattice sites 
$\m=(m_x,m_y)$ the Hamiltonian of the tight-binding
 model of independent electrons  
reads
\begin{equation}
        H=\sum_{\m} \varepsilon_{\m} \left| \m \right\rangle \left\langle \m
\right|
+ \sum_{\BRA \m , \m'\KET } t_{\m ,\m'} \left|\m \right\rangle \left\langle
\m' \right|
\, ,
\label{2}
\end{equation}
where $\BRA \m , \m'\KET $ denotes nearest neighbors.
The site energies $\varepsilon_{\m}$ correspond to the (random)
potential energy and the hopping matrix elements $t_{\m ,\m'}$ to the
kinetic energy. 
The magnetic field is included in the kinetic energy
by the Peierls substitution
\begin{eqnarray}
        t_{\m ,\m'} = 1\;\;\, \quad\quad\quad\;
  &{\rm if}&\;\; m_y=m'_y \;\; {\rm and}
\;\; m'_x=m_x\pm 1  
\label{3a}\\
t_{\m ,\m'} = e^{\pm 2\pi i\alpha m_x} \;\;\; &{\rm if}&\;\;
m_x=m'_x \;\; {\rm and}   
\;\; m'_y=m_y\pm 1   \, . \label{3b}
\end{eqnarray}
Here $\alpha$, restricted to $0< \alpha < 1$,
 is the number of flux quanta $h/e$ per unit cell.
 The characteristic
length
associated to $\alpha$ is the magnetic length $l_B$ defined by
 $\alpha=a^2/2\pi l_B^2$.  Requiring commensurability between the lattice
constant and the magnetic length restricts $\alpha$ to  rational
values
\begin{equation}
        \alpha=p/q\, .
\label{4}
\end{equation}

We impose generalized periodic boundary conditions on the
eigenfunctions in $y$-direction
\begin{equation}
        \psi(m_x,m_y+M_y)= e^{-2\pi i \vartheta}\psi(m_x,m_y)
\label{5}
\end{equation}
and Dirichlet boundary conditions in $x$-direction (cylinder)
\begin{equation}
        \psi(0,m_y)=\psi(M_x+1,m_y)=0
\label{6}
\end{equation}
or (for comparison) periodic boundary conditions in $x$-direction (torus)
\begin{equation}
        \psi(m_x+M_x,m_y)=\psi(m_x,m_y)\, .
\label{7}
\end{equation}
The parameter $\vartheta \in [0,1)$  generalizes the periodicity 
of the wavefunction
 to be periodic ``up to a phase''. 
$\vartheta/M_y$ can, however, also be interpreted as a Bloch quantum number
\cite{Haj87}. Indeed, by
transforming the eigenvalue problem $H\psi=E\psi$
according to the replacement 
\begin{equation}
        \psi(m_x,m_y) \longrightarrow  e^{2\pi
i (\vartheta/M_y)m_y }\psi(m_x,m_y)  
\label{8}
\end{equation}
one arrives at an equivalent problem with strictly periodic boundary
condition in $y$-direction. The transformed hopping
matrix elements  read for $m_x=m'_x$, $m_y=m'_y\pm 1$
\begin{equation}
        t_{\m ,\m'} = e^{\pm 2\pi i\lbrack \alpha m_x + \vartheta/M_y\rbrack} 
 \, .
\label{88}
\end{equation}
A third way to interpret the parameter is by considering the system as
the surface of a cylinder embedded in a 3D space. With $y$ denoting
the coordinate along the circumference of the cylinder, $\vartheta$
describes an Aharonov-Bohm flux (in units of the flux quantum) along
the cylinder axis (cf.~\cite{Haj87}). 

\section{Topology and the Quantum Hall Effect}\label{topology}
We adopt the view of considering a family of Hamiltonians
$H(\vartheta)$ (characterized by Eqs.~(\ref{2},\ref{3a},\ref{88}))
 and wavefunctions that
are periodic in $y$-direction. For each member of that family the
velocity operator $v_y(\vartheta)$
 can be obtained by taking the derivative of $H(\vartheta)$
with respect to the Bloch-momentum $\hbar\vartheta/M_y$,
\begin{equation}
        v_y(\vartheta) =\frac{M_y}{\hbar}\frac{\partial H(\vartheta)}{\partial
\vartheta}\, .
\label{9}
\end{equation}

To model a disordered system we take independent random values for the
site energies $\varepsilon_{\m}$ according to a symmetric box
distribution on the interval $[-V/2,V/2]$. 

For vanishing disorder, $V=0$, and  zero magnetic field, $\alpha=0$,
the Hamiltonian describes an electron moving in a 2D crystal. The
energy band is symmetric around $E=0$ and has the total width $8$. 
On introducing the rational flux $\alpha=p/q$ and adopting toroidal
boundary conditions (Eqs.~(\ref{5}),(\ref{7})  with $\vartheta=0$) 
commensurate with the flux lattice, i.e.~$M_x=ql$ with some integer
$l$,
the spectrum consists of $q$ magnetic subbands (Harper bands)
 separated by energy gaps (cf.~\cite{Hof76} and Fig.~1 of
Ref.~\cite{Sch84}) and the  statements (i) -- (iv) listed in
 the Introduction apply.

For $V\not= 0$ the full 2D problem can be solved by diagonalizing
(numerically) the representing (sparse) Hamiltonian matrix which
yields
the eigenvectors $\psi_l (\m;\vartheta)$ and eigenvalues
$\varepsilon_l(\vartheta)$ labeled by the discrete index
$l$.  In the present work we 
analyze the model by means of energy orbits to be introduced in the following.

Defining the center of mass coordinate 
(in $x$-direction) of a given eigenstate by   
\begin{equation}
        X_{l}(\vartheta):= \sum_{\m} m_x |\psi_l(\m;\vartheta)|^2
\label{10}
\end{equation}
allows us to calculate the Hall conductivity
$\sigma_H=\sigma_{yx}$
by using an appropriate
version of the Kubo formula (cf.~Eq.(4.76))
in\cite{JanB}) which,
 for zero temperature, 
 reads
\begin{equation}
        \sigma_H=-\frac{e^2}{h}\frac{1}{M_x}\int\limits_0^1
        d\vartheta \, 
        \sum_l \Theta \left( \varepsilon_F -
\varepsilon_l(\vartheta)  \right) \frac{\partial
X_l(\vartheta)}{\partial \vartheta}
\label{11}
\end{equation}
where $\varepsilon_F$ is the Fermi energy.
This expression tells that once we know the ``energy-orbits'' 
$(X_l(\vartheta),\varepsilon_l(\vartheta))$ we can
read off the Hall conductivity at a given Fermi energy by determining
the intersection points, $(X^{(i)},\varepsilon^{(i)}=\varepsilon_F)$, 
of the orbits with the line of constant Fermi
energy.
\begin{equation}\label{11B}
        \sigma_H=  \sum_{i}\frac{{\rm sign}(i)\, X^{(i)}
        }{M_x}\left(\frac{e^2}{h}\right)\, .
\end{equation} 
This formula tells that one has to  add
 up the center of mass coordinates of the
intersection points with a signum $\pm$ depending on crossing from
below ($-$) or above ($+$)
and finally divides by the width $M_x$ resulting in the Hall
conductivity (in atomic units). Note that the orbits are (generically) closed, since
level crossings occur with zero probability in a disordered system 
(Wigner-von Neumann theorem \cite{Wig29}). 

Those orbits which connect both edge regions of the cylinder are of
particular importance: they contribute (almost) integer numbers to the
Hall conductivity. 

\medskip
{\em These integer numbers are given by the winding number of
 orbits connecting both edge regions of the cylinder.
 The chirality of the orbits defines the sign of the
integer,
($+$) for clockwise and ($-$) for counterclockwise windings.}
\medskip

 The slight
difference
between the  center of mass coordinate of an edge state $X$ and the
geometrical
edge ($m_x=0,M_x$)  gives rise to corrections. However, they are of
order $(l_B/L_x)$ and furthermore they turn out to be
 artificially caused by using  conductivity instead of conductance
(cf.~the discussion in \cite{Vie90} and Sec.~4.8 of \cite{JanB}).
In our numerical calculations these corrections are still visible for
system sizes of order $100 a$. 
 Equation (\ref{11})
has been analyzed previously in the context of continuous electron systems
on the surface of a cylinder \cite{Haj87,Oht88,Vie90}.
We mention that the very fact of level anti-crossing leads to the presence of
energy gaps (mini gaps) in the regime of edge states. Within these mini gaps
the Hall conductivity vanishes. A similar observation in a two
terminal conductance calculation was denoted in \cite{Skj94} 
as ``cracks in the wings'
of the Hofstadter butterfly. However, these gaps are exponentially
small
in the ratio of magnetic length vs. system size (cf.~\cite{Vie90})
and will be ignored in the following discussions. 

We like to comment further on the topological quantization of the Hall 
conductivity as expressed by the winding numbers of energy orbits.
A  condition for the occurrence of plateaus in the Hall conductivity
is the existence of localized states in the bulk of the system.
 Such states are
characterized
by an exponential decay in space far off their center position.
The decay length is called the localization length $\xi$.
Localized bulk states correspond to orbits degenerated to points since the
sensitivity of the center of mass 
to a change in boundary conditions becomes exponentially
weak in the ratio  $\xi/L_x$. 
If, for a given Fermi energy,  
all bulk states of the system are localized but extended edge states
do exist,
the winding number of these edge states determine the quantized value
of the Hall
conductivity. This discussion is somewhat simplified since localized
bulk states and extended edge states are not degenerate in energy.
Rather, the states at a given energy are hybrids of both types of states (see
Fig.~\ref{fig9}a). 
As already mentioned in the Introduction the topological
quantization in terms of winding numbers is in close analogy
to the topological quantization in terms of Chern numbers occurring in
the torus geometry. The Chern numbers are given by the winding of the 
zeroes of wave functions around the torus when two parameters (Bloch
quantum numbers) for
generalized periodic boundary conditions 
are changed \cite{Koh85,Aro88}: Moving the Bloch quantum numbers over their
entire Brillouin zone (being also a torus) leads to an integer
covering number  on the  torus in real space. This integer
is just the Chern number determining the Hall conductivity. 
Thus, the  Hall conductivity counts  the number of times the zeroes of a wave
function
cover the torus when changing the  boundary conditions. In the
cylinder
geometry the Hall conductivity counts the number of times the center of mass of
a wave function covers the cylinder when the boundary
conditions are changed.
This analogy demonstrates the universality of topological quantization
arguments for the Hall conductivity; they are not restricted to a
particular
geometry although for different geometries the topological
quantization can manifest itself in different objects.

A further indication  for this universality
comes from the observation that the  
distinction between localized and extended states can be based on
topological properties of either model.
In the torus geometry localized states are characterized 
by a vanishing Chern number corresponding to a vanishing covering
number for the zeroes of the wave function.
In the cylinder geometry localized states are characterized by a
vanishing covering number for the center of mass of the wave function.
In a quasi-classical treatment (see e.g.~\cite{Ape85}) localized states
are characterized by closed equipotential lines that do not wind around the
cylinder.

Equation~(\ref{11}) not only helps in demonstrating the quantization
of the Hall conductivity in a regime of localization  but  is also
suitable for discussing the Hall conductivity in the transition regimes
between adjacent plateaus. Here complicated orbits lead to a
strongly fluctuating Hall conductivity, as will be discussed in the
next sections. It is worth mentioning that
similar fluctuations are expected to occur in the torus geometry
whenever the Fermi energy crosses  a band of extended states
(cf.~\cite{Poo87}).

A quantity that is simpler than the Hall conductivity
and usually still allows to distinguish between localized and extended states
is the so-called Thouless number $G$ 
(cf.~\cite{Tho74}) defined  as the ratio of a typical energy shift
$\delta \varepsilon$ 
 within a band and the average
level spacing $\Delta$
\begin{equation}
        G=\frac{\delta \varepsilon}{\Delta}
\label{12}
\end{equation}
This quantity is similar to a dissipative conductance (in atomic
units) since it becomes small compared to $1$ for localized states.
However, as we will see, for the cylinder model there are extended
edge states even in  energy ranges where the bulk states are
localized. Thus, the Thouless number will  show some non-zero
peaks indicating that it is not a suitable criterion for a
bulk state to be extended. Therefore we will focus  in the following
sections on the Hall conductivity, Eq.~(\ref{11}).

\section{Weak disorder}\label{orbitsweak}
\begin{figure}
\begin{center}
\leavevmode
\epsfxsize=7cm
\epsffile[72 185 295 386]{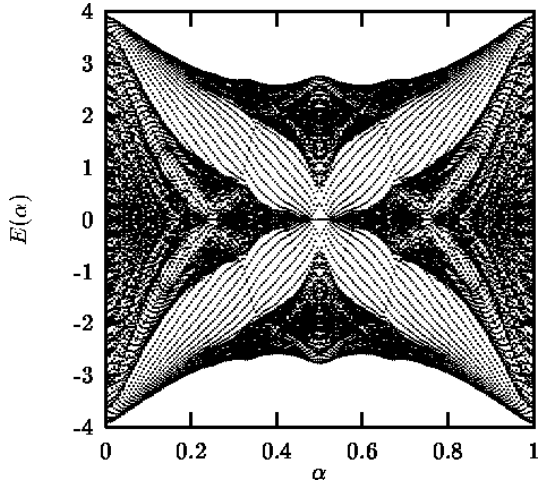}
\end{center}
\caption{Energy spectrum as a function of flux.}
\label{fig1}
\end{figure}

\begin{figure}[!btp]
\begin{center}
\leavevmode
\epsfxsize=7cm
\epsffile[138 505 364 707]{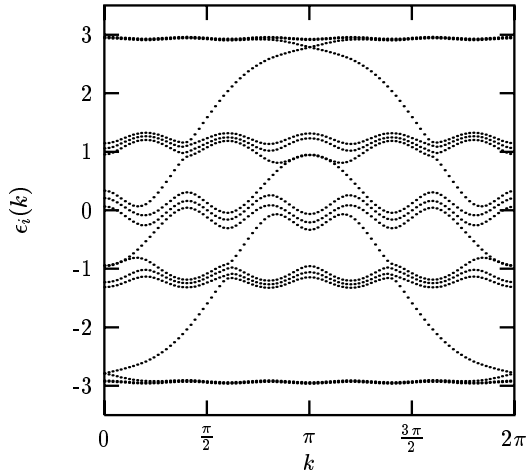}
\end{center}
\caption{Energy branches as a function of wave number for flux $\alpha=1/5$.}
\label{fig2}
\end{figure}

\begin{figure}
\begin{center}
\leavevmode
a)\epsfxsize=7cm\epsffile[136 505 363 707]{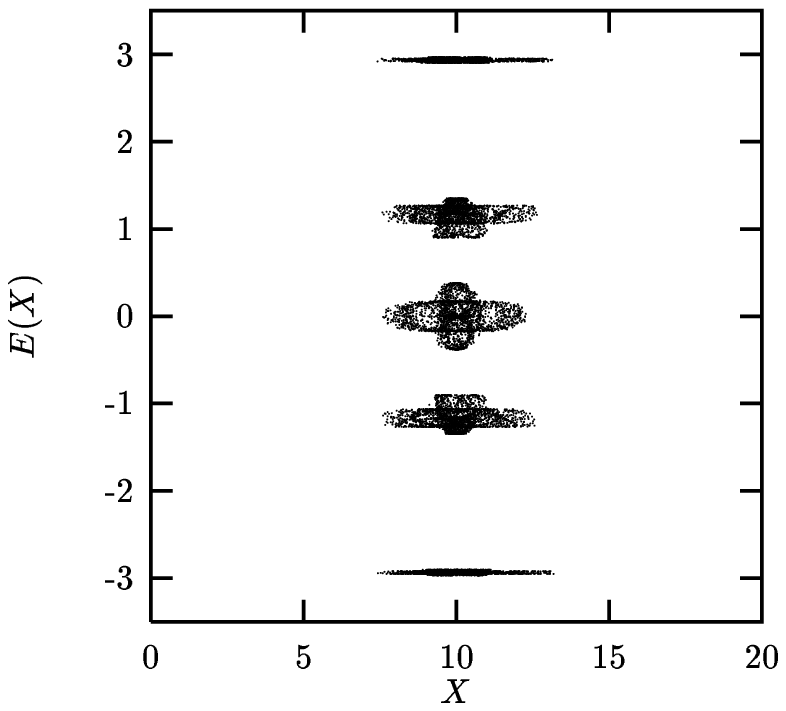}\\
b)\epsfxsize=7cm\epsffile[136 505 363 707]{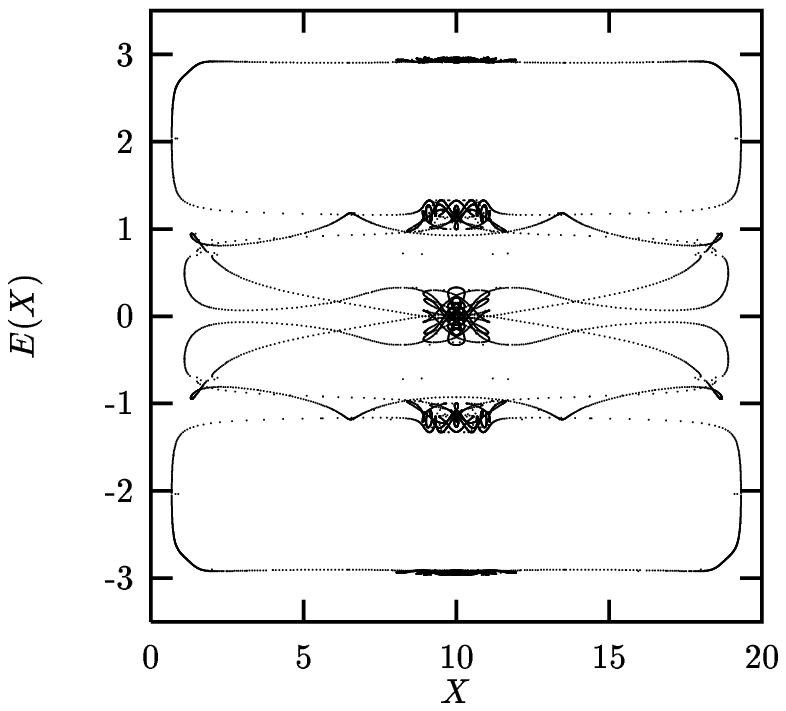}
\end{center}
\caption{Energy orbits for a   system with no disorder
and flux $\alpha=1/5$ for a
torus geometry  (a) and for a cylinder geometry (b).}
\label{fig3}
\end{figure}

We begin our discussion with the limit of vanishing disorder and 
focus on  a significant difference between the torus and the cylinder geometry.
As has been pointed out in \cite{Tou83}
 the difference with respect to the spectrum
is that in the cylinder model  energy
eigenvalues occur in the Harper gaps of the torus model. This can be seen in
 Fig.~\ref{fig1}
where the band structure for the cylinder model
is shown as a function of $\alpha$ (the number of flux
quanta per unit cell).  Regions with dense set of points
correspond to the Harper bands while those  with lines of points
correspond to the gaps (see also Fig.~1 in \cite{Hof76} and Fig.~4 in
\cite{Tou83}). For vanishing disorder the translational invariance in
$y$-direction  yields a wave number $k$ 
labeling
the energy eigenvalues (cf.~\cite{Hof76}). The existence
of energy
branches
within the Harper gaps can be easily observed by plotting 
$\varepsilon_i(k)$.
In Fig.~\ref{fig2} such  plot is shown for $\alpha=1/5$,
$\vartheta=0$
 and $M_x=M_y=17$.
One observes both 
the $5$ Harper bands and some energy branches
within
 the
Harper gaps. To see that the existence of energy eigenvalues in the
Harper gaps is due to edge states it is instructive to plot the energy
eigenvalues versus the corresponding center of mass coordinate (in
$x$-direction) for a number of $\vartheta$ values. The
corresponding plot  for the torus geometry is shown in
part a) of Fig.~\ref{fig3}. This plot not only shows the Harper bands
but also demonstrates that the center of mass, $X$, is distributed within
the bulk of the system. For a  system with no disorder
the wavefunctions are
periodic in $x$-direction \cite{period}. Thus,  $X$  is distributed within a
range smaller than the period. 
 On introducing the cylinder boundary
conditions the picture changes. As can be seen in part b) of
Fig.~\ref{fig3},
the states which have energies within the Harper gaps are concentrated
along the edges (see also Fig.~6 of Ref.~\cite{Skj94}).
 Changing $\vartheta$ from $0$ to $1$ can move an edge
state from one edge to the opposite edge and, finally,  back to the
starting point. On the other hand, the states with energies within the
Harper bands are less effected by the presence of an edge.

The existence of edge states can be qualitatively understood by
considering the Dirichlet boundary conditions together with the
chirality inherent in the phase
of the hopping elements.
For an
electron that starts to propagate close to an edge these items
 lead to a directedness of the most probable path
along the edges. 

\begin{figure}
\begin{center}
\leavevmode
\epsfxsize=7cm
\epsffile[136 505 368 705]{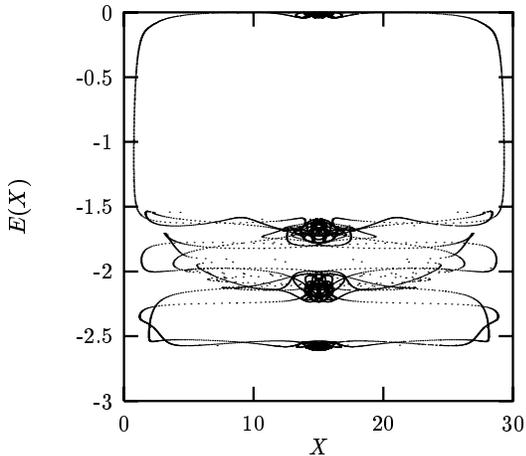}
\end{center}
\caption{Energy orbits for a   cylindrical system with no disorder
and  flux $\alpha=3/7$.}
\label{fig4}
\end{figure}

The energy orbits
$(\varepsilon_l(\vartheta),X_l(\vartheta))$  provide a very helpful
picture of the spectral properties of the cylinder model
 and allow to
calculate the Hall conductivity rather easily, as  discussed
after Eq.~(\ref{11}). 
For example the Hall conductivity for $\alpha=1/5$ is $1,2,-2,-1$ in
the first, second, third and fourth gap (from below),
respectively. These findings are consistent with the Diophantine
equation (\ref{1}). For values $\alpha=1/q$, with $q$ large, the low
lying Harper bands strongly resemble Landau bands of continuous
electron models \cite{Gud88} and the quantization of $\sigma_H$ in
increasing integer steps is no surprise. 
However, in the  lattice
model we can take  a rational value with a dominator different from $1$,
e.g. $\alpha=3/7$. Here the Diophantine equation
 predicts a sequence
of $\sigma_H$ as follows: $\sigma_H= -2,3,1,-1,-3,2$. 
As can be seen in Fig.~\ref{fig4} the  
 result is still valid in the
cylinder model and coincides with  counting the winding numbers
 of orbits connecting edge states. In fact, in all the cases  we
analyzed
the chirality  of the
orbits leads  to the  sign as
predicted by the Diophantine equation.  
Thus, we recover the  result obtained in 
\cite{Hat93} that the  Hall conductivity for
the  cylinder model with vanishing disorder
coincides with the Hall conductivity for the
 torus model with vanishing disorder, as expressed by Eq.~(\ref{1}).
 The point we wanted to make
here was that the   concept of winding numbers of energy orbits
is suitable to determine the Hall conductivity. We will use this concept
further on when increasing successively the  strength of disorder. 

Let us first consider the case of weak disorder. By this we mean
that  the disorder strength $V$ is small as compared
to the strength of the hopping matrix element ( set $1$).
Since for any given matrix size $N$ the number of diagonal elements
and the number of non-zero off-diagonal hopping elements are both of
order $N$, significant changes in global spectral
properties (such as the density of states) occur when $V$ becomes
of order $1$. 
Note, however, that this definition of weak disorder
 does not apply to 
 disorder related  length scales. For example, once the system is large enough
localized states will show up in regimes of  low density of states. Thus,
arbitrarily small amount of disorder can lead to  finite localization
lengths $\xi$. From  this point of view disorder is always strong.

\begin{figure}
\begin{center}
\leavevmode
\epsfxsize=7cm
\epsffile[142 505 368 705]{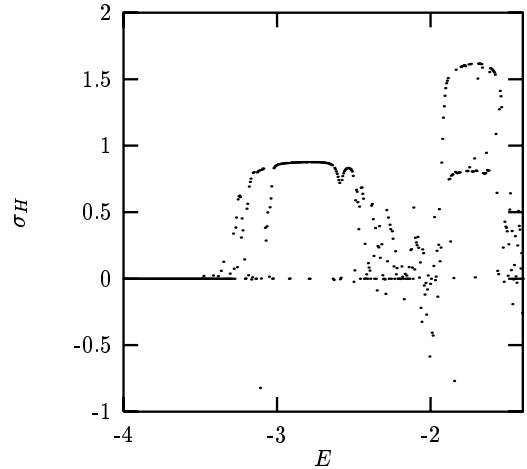}
\end{center}
\caption{Hall conductivity as a function of energy for disorder
strength $V=0.9$, flux $\alpha=1/9$ and system size $M=20$.}
\label{fig5}
\end{figure}

\begin{figure}[!btp]
\begin{center}
\leavevmode
\epsfxsize=7cm
\epsffile[136 505 363 710]{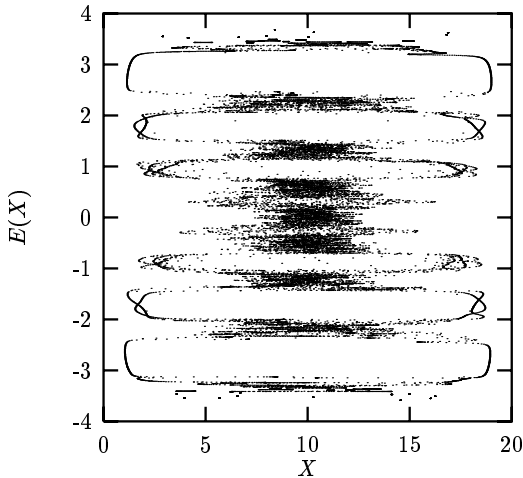}
\end{center}
\caption{Energy orbits for disorder strength $V=0.6$, flux
$\alpha=1/9$ and system size $M=20$}
\label{fig6}
\end{figure}

In the following we will concentrate on a cylinder system with
$\alpha=1/9$ and $M_x=M_y=20$. For $V=0.9$ the Harper bands become
broader but still do not overlap. 
In Fig.~\ref{fig5} the Hall conductivity is
 shown as a function of the Fermi energy
for  $V=0.9$. The quantization of $\sigma_H$ in the gaps remains
valid. However, the behavior of $\sigma_H$ 
within the broadened Harper band is striking: it drops down and
fluctuates around zero.  
A similar behavior can be concluded from the plot of energy orbits for
  $V=0.6$ 
displayed in Fig.~\ref{fig6}.
Again the quantization in the region of bulk states energy gaps
according to the Diophantine equation 
remains valid ($\sigma_H=e^2/h(1,2,3,-3,-2,-3)$) but the variation  of
center coordinates $X$ for extended 
bulk states is restricted to  a range  of
approximately $10$ lattice spacings around the center of the system. 
Therefore, the Hall conductivity will not exceed the value $0.5$ within
the band. The question arises if this behavior will  continue for larger
systems. If it does, the Hall 
conductivity will not interpolate between adjacent plateaus but drop
down in the transition regimes. 
To answer this question we first have to understand the origin of the
drop in the Hall conductivity. Equation (\ref{11}) relates the Hall
conductivity to the range of center coordinates $X$ for a given energy
orbit.
For  systems of no disorder  the amplitude of a wave function is periodic over
the flux lattice. Consequently, the range of $X$ is restricted to
be less than the  period \cite{period}.   As soon as we
introduce disorder we have to ask about the nature of extended bulk
states. It is clear from Eq.~(\ref{11}) that the range of $X$ values
can only increase provided the  amplitudes of an extended wave function are
strongly inhomogeneous. Only then an appropriate shift in the boundary
condition parameter $\vartheta$ can lead to a drastic shift in the
position $X$ of the center of mass.

\begin{figure}
\begin{center}
\leavevmode
a)\rotate[r]{\epsfysize=6cm\epsffile[221 119 514 673]{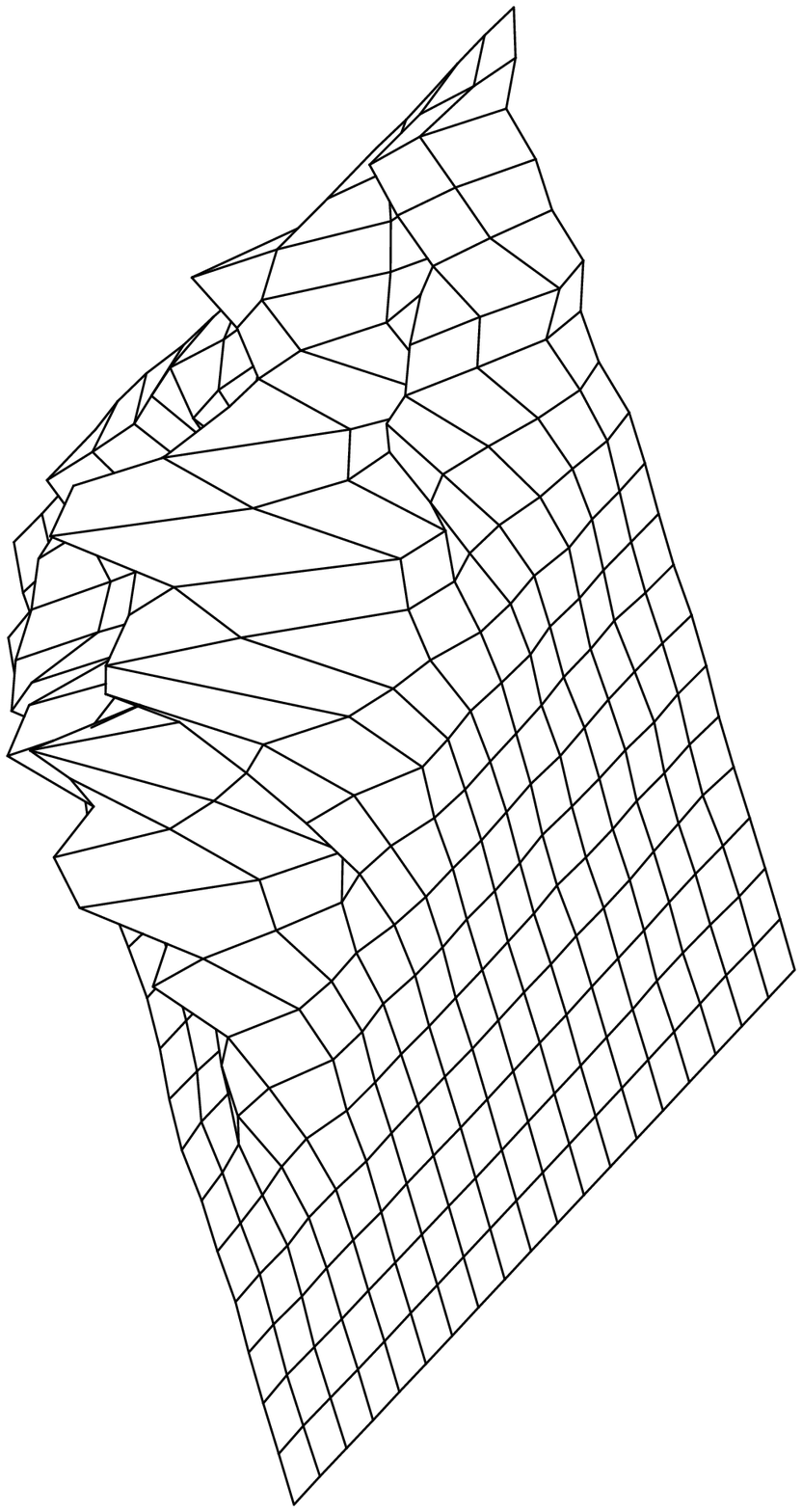}}\\
b)\rotate[r]{\epsfysize=6cm\epsffile[133 119 514 673]{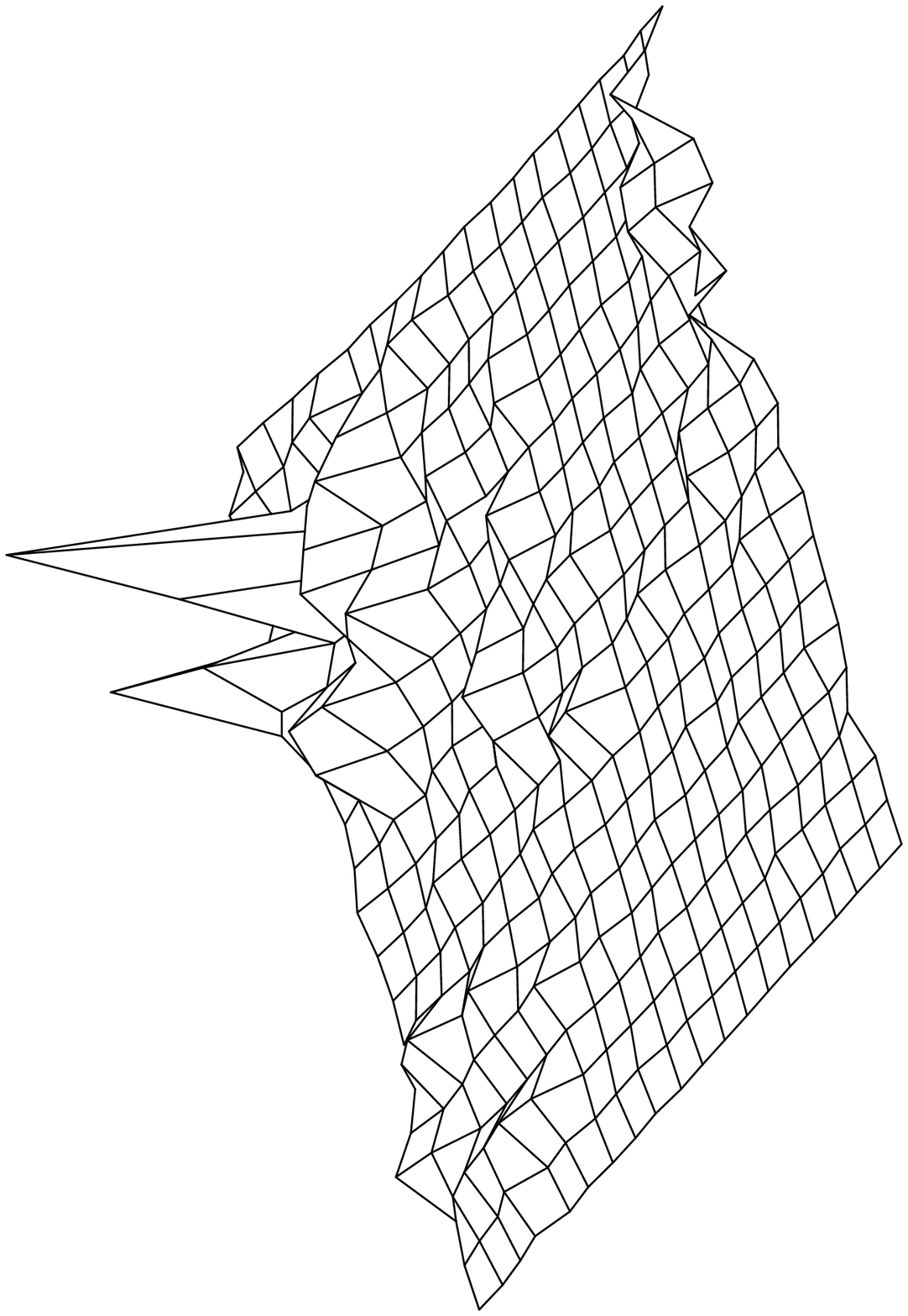}}
\end{center}
\caption{Density plot of two wave functions with energy
in the tail of the first
Harper band (a) and within the center of the second Harper band for
the same system parameters as in Fig.~6.}
\label{fig7}
\end{figure}

The localization length $\xi$ 
of a wavefunction is the relevant length scale 
 introduced  by  disorder.
It is expected that for sufficiently large  system sizes extended
states can only exist in the center of the broadened Harper bands.
All other bulk states are localized and do not contribute to
$\sigma_H$. The bulk extended states are characterized by their
localization length $\xi$ being larger than the actual system size. In
the thermodynamic limit    
$M=M_x=M_y\to\infty$ the spectral width of extended bulk states is
expected to shrink to zero as $M^{-1/\nu}$ where $\nu$ is the critical
exponent of the localization length (cf.~\cite{Huc94}).
 As long as  the system size is so small that we hardly find localized
states,
even in band tails, the system size has to be considered as
``microscopic''.
 The plot of the squared amplitudes (density plot) of two wave functions
corresponding to the tail of the first band and to the center of the
second band of bulk  
states are shown in
Fig.~\ref{fig7}a) and Fig.~\ref{fig7}b), respectively.  The first
state shows a tendency to localize, however the localization length is
of the order of the system size. The second state is delocalized and
already displays some inhomogeneous amplitude fluctuations. However,
for the system size of $M=20$ the center of mass cannot be shifted
over the entire scale of $x$-coordinates. Nevertheless, the second
state already indicates that the extended bulk states can become
spatially inhomogeneous.

\section{Strong disorder}\label{orbitsstrong}
\begin{figure}
\begin{center}
\leavevmode
\epsfxsize=7cm
\epsffile[138 505 366 700]{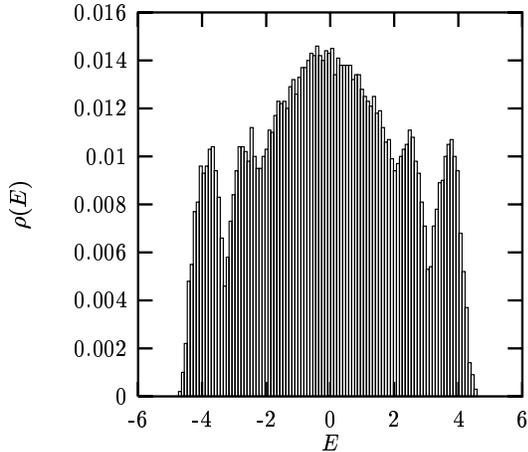}
\end{center}
\caption{Density of states for flux $\alpha=1/9$ and disorder strength
$V=2$ and system size $M=100$.} 
\label{fig8}
\end{figure}

To see the influence of stronger disorder on the Hall conductivity
one can either consider larger systems with fixed disorder strength $V$
or  increase the disorder strength $V$. In both cases 
 a reasonable amount
of localized states show up in the low density of states regime. We
 changed the disorder strength and system size  for  $\alpha=1/9$
to $V=2$ and $M=100$, respectively. The influence of the increased
disorder strength can be observed in the density of states which is
shown in Fig.~\ref{fig8}. The Harper bands strongly overlap and the 
resolution of only the first and the last Harper gap is possible.
These gaps are expected to be filled with localized states and form
mobility gaps rather than spectral gaps.
We will concentrate on the wave function properties for energies
situated within the lowest mobility gap and in the center of the second
Harper band.

\begin{figure}
\begin{center}
\leavevmode
a)\rotate[r]{\epsfysize=6cm\epsffile[181 203 540 756]{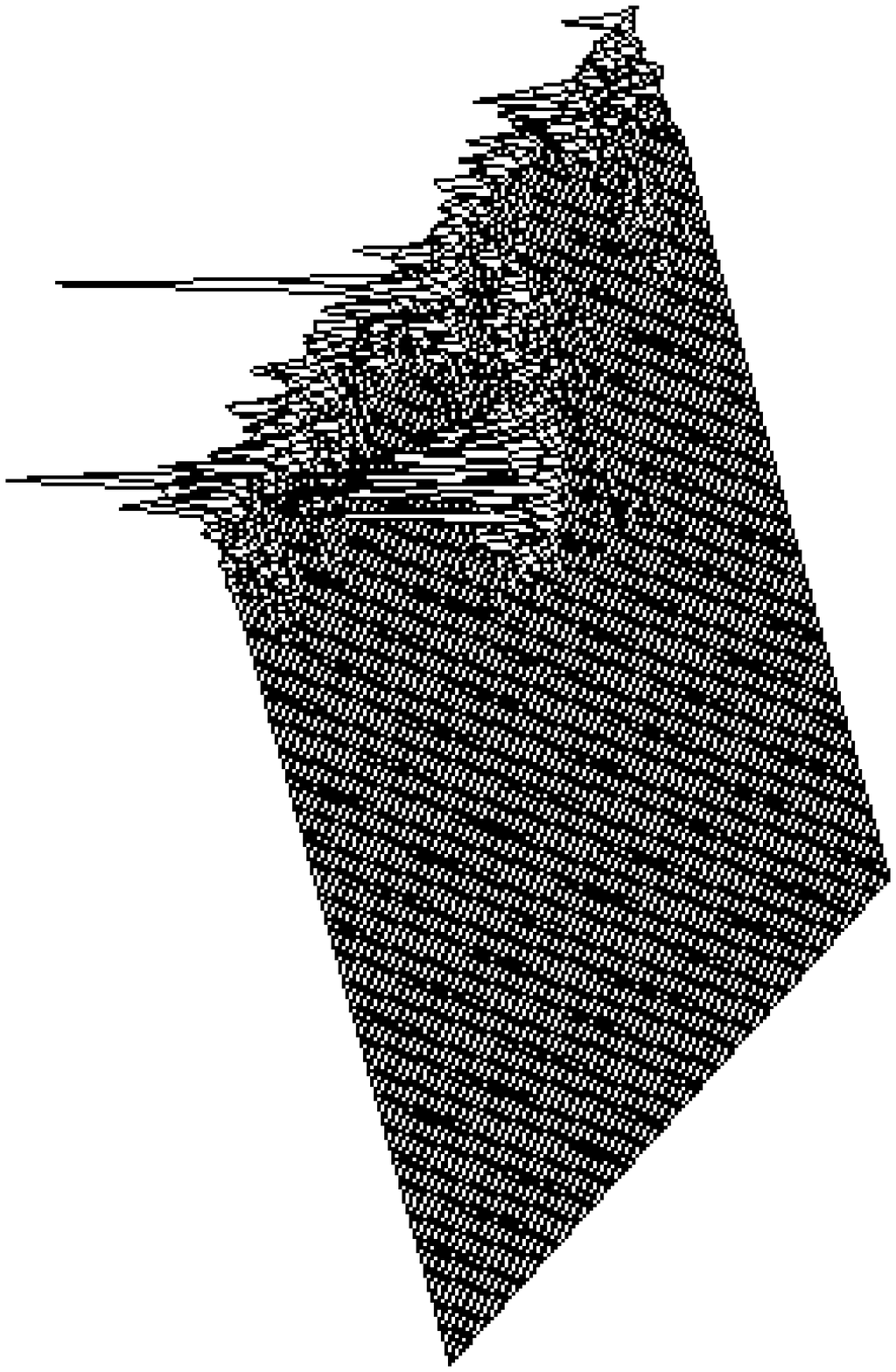}}\\
b)\epsfxsize=6cm\epsffile[136 505 371 695]{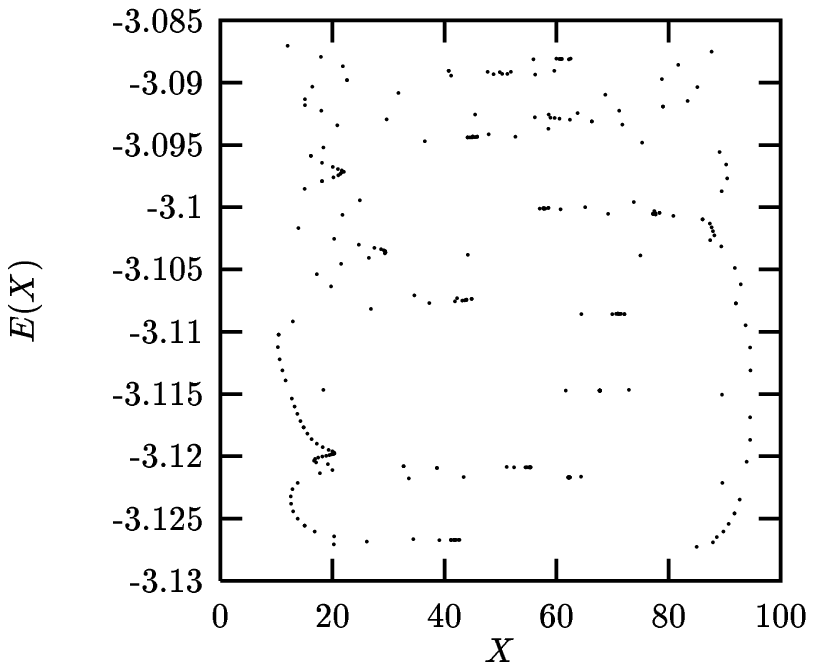}
\end{center}
\caption{Density plot of a wave function with energy in a mobility gap
(a) and the corresponding energy orbits (b). The system parameters
are the same as in Fig.~8. }
\label{fig9}
\end{figure}

We find that within the gap all bulk states are now localized and only extended
edge states appear, which meets the expectations formulated above.
This can be  seen in Fig.~\ref{fig9}a) where the density
plot of a wave function corresponding to a gap energy is displayed.
The plot reflects that the wave function has a bulk contribution
concentrated on an area with diameter of approximately $16$ lattice
spacings
and an edge contribution which is extended along one edge in
$y$-direction.
In contrast to the edge contribution the bulk contribution
is almost insensitive to a change in the
boundary conditions. For a certain value of  the parameter
$\vartheta$ the edge contribution will sweep to the opposite edge,
resulting in a quantized Hall effect for Fermi energies situated in
the mobility gap. The corresponding energy orbits are shown in
Fig.~\ref{fig9}b)
where the movement  of edge states can be seen. 

\begin{figure}
\begin{center}
\leavevmode
a)\rotate[r]{\epsfysize=6cm\epsffile[209 203 487 757]{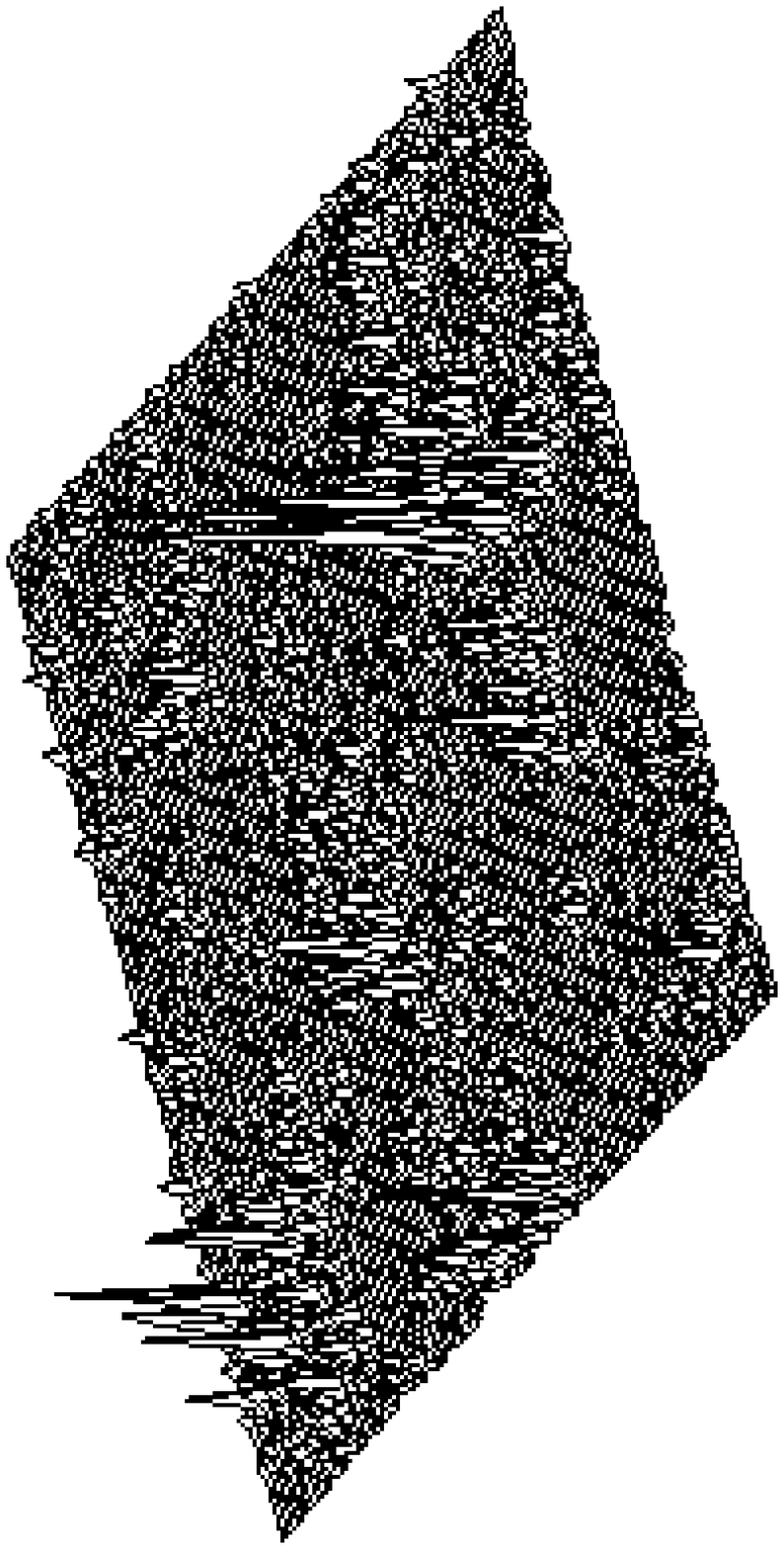}}\\
b)\epsfxsize=6cm\epsffile[136 505 371 695]{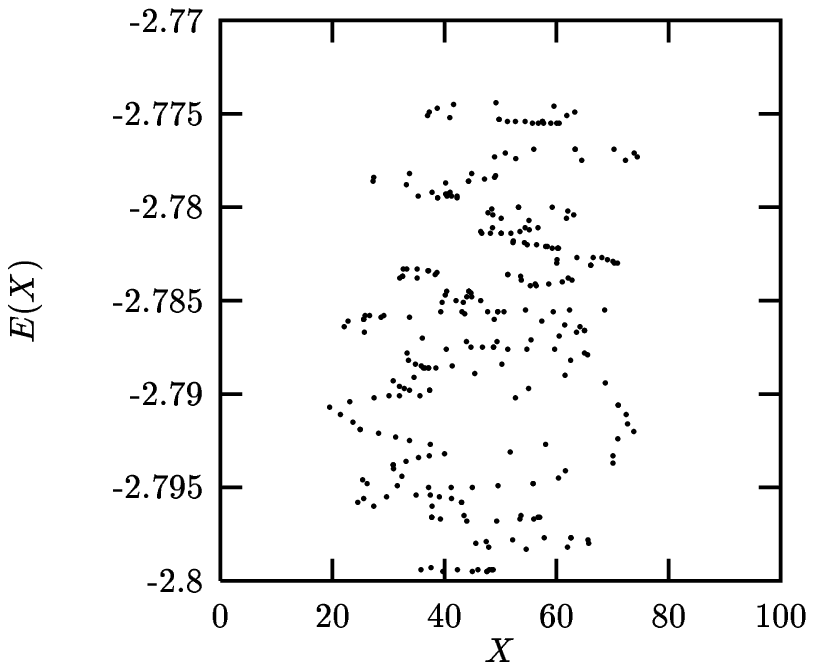}
\end{center}
\caption{Density plot of a wave function with energy in the center of
the second Harper band
(a) and the corresponding energy orbits (b). The system parameters
are the same as in Fig.~9.}
\label{fig10}
\end{figure}

The density plot of Fig.~\ref{fig10}a) shows a state within  the
center of the
second Harper band. It is a typical example of a critical eigenstate
which shows strong spatial amplitude fluctuations. As a consequence
the position of its center of mass $X$ can be shifted over a large
range of values by an appropriate shift in the parameter $\vartheta$,
as seen in Fig.~\ref{fig10}b).  By this observation we are led to
conjecture that 
the corresponding Hall conductivity 
will
interpolate between adjacent plateau values when  increasing further the
system size. Still, the Hall conductivity shows strong (mesoscopic)
fluctuations of ${\cal O}(e^2/h)$ related to the fluctuations of the
center of mass coordinate $X$.     
 For this  conclusion it is essential that extended
bulk states show strong spatial fluctuations.
Such fluctuations are a generic feature of electron states at the
localization-delocalization transition. In addition, these fluctuations
appear on all length scales  between microscopic scales and
the (diverging) macroscopic localization length and consequently
 the corresponding
density plot represents a multifractal (cf.~Chap.12 of \cite{JanB}).
An important feature of critical states is that the typical density,
defined as a geometric mean,
$P_{\rm typ}=\exp \left\langle \ln |\psi|^2\right\rangle$ scales in a
universal manner with the system size $M$
\begin{equation}
        P_{\rm typ} \propto M^{-\alpha_0}\label{ALPHA}
\end{equation}
where $\alpha_0$ is a critical exponent characteristic of the
localization-delocalization transition at hand. The fact that
$\alpha_0$ is larger than the space dimension indicates the difference
between average and typical values of densities and is a criterion for
multifractality. 

In the work by Huckestein {\em et al.} \cite{Huc92}
the critical eigenstates of the lowest Harper band for $\alpha=1/8$
have been analyzed for a torus geometry. They found the
multifractal exponent $\alpha_0$ to be consistent with values reported
for a number of models related to the quantum Hall effect
\cite{Pok91Kle95}. 
A recent calculation \cite{Fre96} yields a value of
$\alpha_0=2.28 \pm 0.02$. For 
 a few of the critical states in the two lowest Harper bands
($\alpha=1/9$, $V=2$, $M=100$) we calculated the exponent
$\alpha_0$ and found values of $\alpha_0$
compatible with those reported previously. This indicates that in the
center of the Harper bands localization-delocalization transitions
occur in accordance with  the quantum Hall universality class.

\section{Conclusion}\label{conclusion}
We presented numerical calculations of energy eigenvalues and wave functions
for disordered lattice electron systems on a cylinder surface in the
presence of  rational magnetic flux. We analyzed the Hall conductivity
by means of  closed orbits in the
 energy
 vs. center of mass plane. The orbits are
 parameterized by
 an Aharonov-Bohm flux along the cylinder axis.
 The spectrum is found to consist of disorder broadened Harper bands
with Harper gaps filled by localized bulk states and/or delocalized
 edge states.
In the regime of these bulk mobility gaps the Hall conductivity is
quantized and given by the Diophantine equation, Eq.~(\ref{1}),
 originally derived for
clean systems with torus geometry.

In close analogy to systems with torus geometry the quantized Hall
 conductivity for systems with cylinder geometry is shown to be
 related to topological quantum numbers. Whereas for the torus these
 are Chern numbers, they are the winding numbers of the above defined
 orbits in the case of the cylinder. Thus, concerning the often discussed
 ``topological explanation'' of the quantum Hall effect both
 geometries are shown to be essentially  equivalent.

 Within the broadened Harper bands
a regime of extended bulk states exist. 
Regarding   the drop in the Hall conductivity
for very small and weakly disordered  systems, we do not expect that this
drop will occur in a real multi-terminal measurement. 
Let us stress
that the Hall conductivity {\em assumes} a uniform electric field
within the system while a true conductance must be based on the
self-consistent field within the system. As discussed in Ref.~\cite{JanB} the
difference
between conductance  and conductivity 
is negligeble (${\cal O}(l_B/L)$) 
for discussing the plateau region, however this is not
the case  in the region of bulk-extended states. 
These bulk states are, for
sufficiently
large
enough systems, multifractal   with universal
characteristics of critical states in quantum Hall systems. The
multifractality is directly connected to the mesoscopic fluctuations
of the Hall conductivity within the regime of critical states; the
sensitivity of the center of mass to the Aharonov-Bohm  flux leads to drastic
fluctuations of the Hall conductivity. These fluctuations are of order
$e^2/h$
and the distribution of the Hall conductivity in the transition regime
between Hall plateaus deserves further studies. Although the very fact of
mesoscopic fluctuations in the conductivity points to a similar effect
for the conductance, a direct quantitative comparison is not possible. 
 The work by  Aldea {\em et al.} \cite{Ald96},
where
multi-terminal conductances were calculated (for systems of 
$\approx 600$ lattice
sites),   gives support to
the expectation that a true multi-terminal Hall conductance fluctuates
between subsequent plateaus. 
 The experimentally  observed
flat distribution \cite{Cob96} of two-terminal conductances
in the transition regime is compatible with the fluctuations we
found in our calculations of the Hall conductivity. They have  been 
reproduced for the  two-terminal conductance 
within a (one Landau band)
network model of a  quantum Hall system \cite{Cho96}. 
 In our model similar
fluctuations correspond to  the strong amplitude fluctuations in
critical 
(multifractal) eigenstates.

\bigskip
\bigskip

\begin{center}
  {\bf Acknowledgment}
\end{center}

This work was performed within the research program of the
Sonderforschungsbereich 341 of the Deutsche Forschungsgemeinschaft.


\begin{thebibliography}{99}

\bibitem[*]{goetze}
Dedicated to Professor
Wolfgang G\"otze on the occasion of his 60th birthday

\bibitem{PraGirv} R.E. Prange, S. Girvin 
(Eds.), The Quantum Hall Effect,
Springer,
New York (1990).
 
\bibitem{JanB}  M.  Janssen, O.  Viehweger, U.  Fastenrath, J.
Hajdu, 
Introduction to the Theory of the Integer Quantum Hall Effect,
VCH, Weinheim (1994).

\bibitem{Hof76} D.R. Hofstadter, Phys. Rev. B{\bf 14}, 2239 (1976).

\bibitem{Exp} T. Schloesser, K. Ensslin, J.P. Kotthaus, M. Holland,
Europhys. Lett. {\bf 33}, 683 (1996).


\bibitem{Skj94} J. Skj{\aa}nes, E.H. Hauge, G. Sch\"on, Phys. Rev. B{\bf 50}, 
 8636 (1994).

\bibitem{Ald96} A. Aldea, P. Gartner, A. Manolescu, M. Nita, unpublished
(1996).

\bibitem{Tho82} D.J. Thouless, M. Kohmoto, M.P. Nightingale, M. den Nijs,
Phys. Rev. Lett. {\bf 49}, 405 (1982).


\bibitem{Tou83} R. Rammal, G. Toulouse, M. T. Jaekel, B. I. Halperin,
Phys. Rev. B {\bf 27}, 5142 (1983).

 
\bibitem{Hat93} Y. Hatsugai, Phys. Rev. B{\bf 48}, 11851 (1993);
Phys. Rev. Lett. {\bf 71}, 3697 (1993).

\bibitem{Str94} P. St{\v r}eda, J. Ku{\v c}era,
 D. Pfannkuche, R.R. Gerhardts,
A.H. MacDonald, Phys. Rev. B {\bf 50}, 11955 (1994).

\bibitem{Huc94} B. Huckestein, Rev. Mod. Phys. {\bf 67}, 357 (1995).

\bibitem{Sch84} L. Schweitzer, B. Kramer, A. MacKinnon, J. Phys. C {\bf
17}, 4111 (1984).

\bibitem{Aok84} H. Aoki, J. Phys. C{\bf 18}, L67 (1984).

\bibitem{Haj87} J. Hajdu, M. Janssen, O. Viehweger, Z. Phys. B{\bf
66}, 433 (1987).

\bibitem{Koh85} M. Kohmoto, Ann. Phys. (N.Y.) {\bf 160}, 343 (1985).

\bibitem{Aro88} D. P. Arovas, R. N. Bhatt, F. D. M. Haldane,
P. B. Littlewood, R. Rammal, Phys. Rev. Lett. {\bf 60}, 619 (1988).



\bibitem{Wig29} E. Wigner, J. von Neumann, Z. Phys. {\bf 30}, 467 (1929).
 

\bibitem{Vie90} O. Viehweger, W. Pook, M. Jan{\ss}en, J. Hajdu,
Z. Phys. B {\bf 78},  11 (1990).

\bibitem{Oht88} T. Ohtsuki, Y. Ono, Solid State Commun. {\bf 65}, 403 (1988).


\bibitem{Ape85} S. M. Apenko, Y. E. Lozovik, Sov. Phys. JETP {\bf 62(2)},
328 (1985).

\bibitem{Poo87} W. Pook, J. Hajdu, Z. Phys. B {\bf 66}, 427 (1987).


\bibitem{Tho74} D. J. Thouless, Phys. Rep. {\bf 13 C}, 93 (1974).

\bibitem{period} The period is typically  equal to $q$.
Due to degeneracies larger periods equal to an integer multiple of $q$ can
occur.



\bibitem{Gud88} V. Gudmundsson, R.R. Gerhardts, R. Johnston,
L. Schweitzer, Z. Phys. B{\bf 70}, 453 (1988).






\bibitem{Huc92}
B. Huckestein, B. Kramer, 
L. Schweitzer, Surface Sciences {\bf 263,} 125 (1992).

\bibitem{Pok91Kle95} W. Pook, M. Janssen, Z. Phys. B {\bf 82,} 295 (1991);
R. Klesse, M. Metzler, Europhys. Lett. {\bf 32,} 229 (1995).
 
\bibitem{Fre96} P. Freche, unpublished (1996).

\bibitem{Cob96} D. H. Cobden, E. Kogan, unpublished, cond-mat/9606114 (1996).

\bibitem{Cho96} S. Cho, P. A. Fisher, unpublished, cond-mat/9609048 (1996).

\end{thebibliography}
\end{document}